\documentclass[pre,twocolumn,showpacs,preprintnumbers,floatfix]{revtex4}

\usepackage{graphicx,citesort,amsmath,amssymb}

\begin{document}

\title{A cluster mode--coupling approach to weak gelation in attractive
colloids}

\author{$^{1,2}$K. Kroy}
\author{$^{2}$M. E. Cates}
\author{$^{2}$W. C. K. Poon}
\affiliation{$^{1}$Hahn--Meitner Institut, Glienicker Str.~100, 14109
Berlin, Germany}
\affiliation{$^{2}$ School of Physics, The University of
Edinburgh, Edinburgh EH9 3JZ, United Kingdom}

\begin{abstract}
Mode--coupling theory (MCT) predicts arrest of colloids in terms of
their volume fraction, and the range and depth of the interparticle
attraction. We discuss how effective values of these parameters evolve
under cluster aggregation. We argue that weak gelation in colloids can
be idealized as a two--stage ergodicity breaking: first at short
scales (approximated by the bare MCT) and then at larger scales
(governed by MCT applied to clusters). The competition between arrest
and phase separation is considered in relation to recent
experiments. We predict a long--lived `semi--ergodic' phase of mobile
clusters, showing logarithmic relaxation close to the gel line.
\end{abstract}
\maketitle

Hard--sphere colloids with short--range attractions can undergo several
types of arrest. At high densities they show two distinct glass
transitions (repulsion--driven and attraction--driven), with a
re--entrant dependence on attraction strength
\cite{pham-etal:2002}. This scenario was first predicted by mode
coupling theory (MCT)
\cite{bergenholtz-fuchs:99,fabbian-etal:99,dawson-etal:2001}, and depends on
both the attraction range $\delta$ (in units of particle diameter) and
well--depth $\varepsilon$ (in units of $k_BT$). MCT is remarkably
successful, at least for large volume fractions $\phi \gtrsim 0.4$.

At lower volume fractions, however, there is no comparable theoretical
framework. Yet `weak gelation', in which
bonding is strong but not so strong as to be irreversible, can lead to
nonergodic soft solids, of nonzero static elastic modulus, at volume
fractions of just a few percent \cite{poon-haw:97}.  It might be
argued that a finite modulus requires a percolating network of bonds
whose lifetime exceeds that of the experiment.  However this is
simplistic: as shown by the case of repulsive glasses, a finite
modulus can arise with no bonding at all.  We argue here that the
rigidity of weak gels arises not from bond percolation but from
kinetic ergodicity breaking \cite{bergenholtz-fuchs-voigtmann:2000},
just as it does in glass formation.  This suggests that an MCT--like
approach to weak gelation could be fruitful.

MCT takes its structural input from equilibrium liquid state theory;
it cannot address states of arrest where this structure is strongly
perturbed \cite{foot1}. This matters relatively little at $\phi
\gtrsim 0.4$, where each particle interacts with many others, and not
much room is left for structural development upon a quench. But more
severe consequences must be expected at low volume fractions where
strongly nonuniform, ramified gels arise. Here the pathway to complete
nonergodicity (starting from a homogenized fluid sample, say) must
involve a nontrivial episode of structure formation, akin to
irreversible cluster aggregation (ICA).  Such kinetics certainly
dominates for irreversible (``strong'') gelation ($\varepsilon^{-1}
=0$), where particles aggregate on contact into clusters, with various
kinetic universality classes \cite{vicsek:92}. Relative simplicity is
restored at low $\phi$ thanks to the invariance of this aggregation
process under coarse graining in the (ordered) limit $\delta$,
$\varepsilon^{-1}\to0$ and $\phi\to0$.  This scaling limit is
controlled by an ICA `fixed point', where details of the short--ranged
attraction are irrelevant. The resulting fractal clusters grow
indefinitely only if $\phi = 0$; for $\phi > 0$ they eventually form a
percolating gel of locally ICA--like structure \cite{hasmy-jullien:96}.

In this Letter, we explore how ICA might connect to weak gelation
through a unified (albeit speculative) MCT--based scenario of
colloidal arrest.  We consider the changes in the system parameters
($\phi, \delta, \varepsilon^{-1}$) upon coarse graining in the
vicinity of the ICA fixed point. This leads to a schematic description
of the suspension in terms of an effective theory for a dense liquid
of `renormalized particles' or coarse--grained clusters, for which
recent simulations \cite{delgado-etal:2003} seem to provide some direct
evidence. Applying MCT to this theory, we obtain a new condition for
arrest {\em of the clusters}, and identify the arrested state with the
weak gel phase. If this condition is not met, one has instead a fluid
of clusters (or `cluster phase' \cite{segre-etal:2001}) that is
ergodic at large scales. We allow for bond--breaking and reconnection
at the scale of entire clusters, but not reconstruction at shorter
scales. We return to this later, where we also address the potentially
complex interplay between weak gelation and phase separation.
 
\begin{figure}[t]		
\begin{center}
\includegraphics[width=\columnwidth]{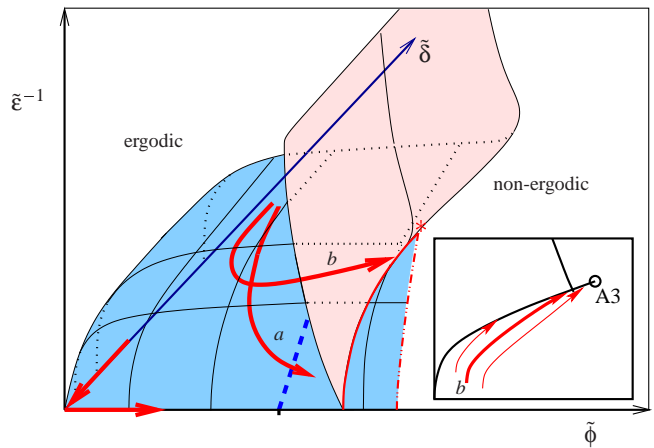}
\end{center}
\caption{Schematic parameter flows (arrows) near the ICA fixed point
(at the origin, bottom left). Flow ($a$) within the lower coordinate
plane ($\varepsilon^{-1} = 0$) gives percolation of ICA clusters on
passing through a non--equilibrium percolation line (heavy dashed).
For $f(N)\geq0$, trajectories starting from finite $\varepsilon^{-1}$
flow {\em upwards} to smaller $\tilde\varepsilon$ and towards
larger $\tilde\phi$ and smaller $\tilde\delta$. Those that arrive on
the domed dark surface result in semi--ergodic cluster phases.  The
marginal gel trajectory ($b$) ends at the seam where the repulsive
(light surface) and attractive (dark surface) branches of the MCT
transtion meet, close to a line of $A_3$ singularities (dash--dotted)
or/and an $A_4$ singularity ($*$). \emph{Inset:} $\tilde\phi-\tilde
\varepsilon^{-1}-$projection of trajectories leading to cluster or
gel phase (thin) and marginal gel ($b$).}
\label{fig:flow}
\end{figure}

Within our scenario, weak colloidal gelation emerges as a double
ergodicity breaking: once when the original suspension becomes
kinetically unstable against aggregation, and a second time when the
fluid of clusters arrests. We suppose that the criterion for the
initial instability is given by bare MCT
\cite{bergenholtz-fuchs-voigtmann:2000}. We assume $1\ll \varepsilon <
\infty$: bonds can break with a small but finite
probability. Initially this has almost no effect on the aggregation,
which flows towards the ICA fixed point. But as clusters grow, the
chances of fragmentation increase. The ICA fixed point is unstable and
the sytem eventually flows away from it. The flow takes place in a
parameter space comprising effective system variables
$(\tilde\phi,\tilde\delta,\tilde\varepsilon^{-1})$
(Fig.~\ref{fig:flow}).

These coarse--grained parameters describe the density and interaction
of aggregated clusters.  We write:
\begin{align}
\tilde\phi & = \phi\; N^{3/d_c-1} \label{eq:n} \\ 
\tilde\delta &= \; \delta \;
N^{-\nu}
\label{eq:d} \; \\
\tilde\varepsilon &=
\varepsilon - \chi\ln N - f(N) \label{eq:e} \;
\end{align}
and call the combination of MCT with Eqs.(\ref{eq:n}--\ref{eq:e})
`cluster mode--coupling theory' (CMCT).  $N$ is the number of
particles in a typical cluster, which is a parametric label along flow
trajectories, each of which is indexed by values of
$(\phi,\delta,\varepsilon^{-1})$ in the initial state after the
quench. In Eqs.(\ref{eq:n},\ref{eq:d}), evolution of the
effective volume fraction of clusters (viewed as
quasi--spherical, smooth objects) involves a scaling index $d_c$, and
that of the effective relative range $\delta$ of their attraction
another positive exponent $\nu$.

More subtle is the evolution of the bond energy $\varepsilon$ in
Eq.(\ref{eq:e}).  Since the ICA fixed point leads only to
singly--connected bonding, there is no power of $N$ multiplying the
short--range attractive part of the bond energy. However,
Eq.(\ref{eq:e}) accounts for bond {\em breaking} through a
renormalization of the cluster--cluster attraction $\varepsilon$ by a
breaking entropy, which we write as the logarithm of a breaking
degeneracy $N^{\chi}\leq N$. This acknowledges that a ramified cluster
falls apart when any of the $N^{\chi}$ bonds along its backbone is
broken (with probability $\sim e^{-\varepsilon}$).  Eq.(\ref{eq:e})
also includes a term $f(N)$ arising from additional interactions. For
purely short range attractions this term should vanish, but it allows
us to address also the case of a weak long range repulsion, such as
has recently been shown to arise from poorly screened dissociated
surface charges even for supposedly neutral colloids in organic
solvents \cite{yethiraj-blaaderen:2003}.  Depending on the strength
and range of the repulsion, the resulting $f(N)$ may dominate the
logarithmic term in Eq.(\ref{eq:e}) without having much effect on the
equilibrium phase diagram \cite{barrier}.

Initially, all three exponents introduced in the CMCT equations are
properties of the ICA fixed point (typically $d_c\approx2$,
$\nu\approx \chi\approx 1/d_c$, in three dimensions
\cite{vicsek:92}). However, they will start evolving once the flow
moves away from it, as the internal structure of clusters evolves.  An
explicit treatment of this drift is not attempted here: we treat
$d_c,\nu,\chi$ as negotiable constants.

\begin{figure}[t]		
\begin{center}
\includegraphics[width=\columnwidth]{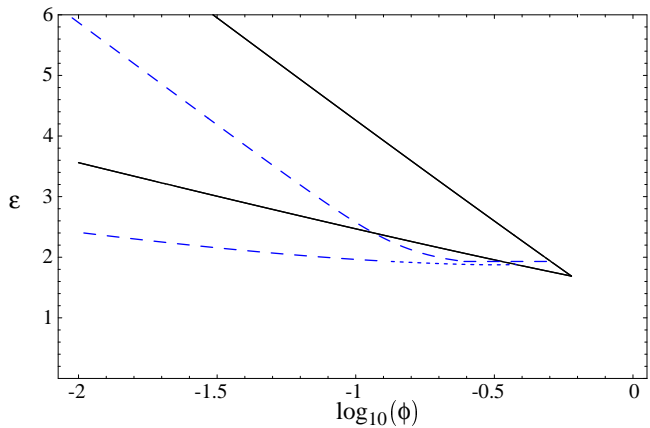}
\end{center}
\caption{Kinetic phase diagram predicted from CMCT with initial range
of square--well attraction $\delta = 0.1$.  Exponents are chosen for
definiteness as $d_c=2,\chi = \nu = 1/2$ and $f(N) \equiv 0$.  Bare MCT
(in the low--$\phi$ approximation
\cite{bergenholtz-fuchs-voigtmann:2000}) is used to estimate the onset
of a clustering instability (lower solid line). The CMCT line divides
cluster fluids from weak gels (upper solid line). Results allowing
also for phase separation, using the indicative model of
Fig.~\ref{fig:gelgas} give shifted phase boundaries (dashed
lines). (The part of the equilibrium binodal above the AB tieline in
Fig.~\ref{fig:gelgas} is also shown, dotted.)}
\label{fig:segrelike}
\end{figure}

The resulting CMCT flow within the renormalized parameter space
$(\tilde\phi,\tilde\delta,\tilde\varepsilon^{-1})$ is sketched in
Fig.~\ref{fig:flow}.  The flow is bounded by the MCT transition
surface in this space, which divides ergodic (upper left) from
nonergodic regions.  This surface
\cite{bergenholtz-fuchs:99,fabbian-etal:99,dawson-etal:2001} comprises
a dome joined to a sheet. Below the dome lie attractive glasses and to
the right of the sheet, repulsive ones.  Note the presence of a seam
$\tilde\phi_c$, on the transition surface, along the line where the
sheet meets the dome.  Consider a trajectory of increasing $N$ which
rises up through the dome, left of the seam
($\tilde\phi<\tilde\phi_c$). At this point, aggregation ceases,
because the clusters' effective parameter values now correspond to an
ergodic phase. A semi--ergodic fluid of finite clusters can be expected
whenever the initial parameter values belong to such a
trajectory. (This requires small $\delta$.)  If a series of such
cluster fluids is now created by increasing $\phi$ at fixed
$\varepsilon$ (or vice versa), then the last of these to enter an
ergodic phase will do so just beside the seam
($\tilde\phi\to\tilde\phi_c^-$). Accordingly the first fully
nonergodic phase formed has a trajectory that just meets this seam;
beyond lies a {\em repulsive} glass, so that aggregation
ceases. Unlike the cluster fluid, the resulting {\em marginal gel} has
finite modulus --- as do the non--marginal gels arising at larger
$\phi$ or $\varepsilon$.  This not because of percolation, which for
reversible bonding is not relevant, but because the CMCT flow has led
to a ``cluster glass'' which we identify with the weak gel phase.  A
typical phase diagram predicted by this approach is shown in
Fig.~\ref{fig:segrelike}.

Note that for moderate $\tilde\delta$, the seam is not too far from
the line of $A_3$ singularities, which is associated with a
distinctive arrest scenario including a large exponent for the
divergence of the structural relaxation time, and strongly stretched
exponential or even logarithmic decay of the density correlations
\cite{dawson-etal:2001,goetze-sperl:2002,sciortino-tartaglia-zaccarelli:tbp}.
Marginal gels could thus be influenced by these singularities even at
low volume fractions, although the details of this depend on the
initial range $\delta$ and on the exponent $\nu$.

CMCT clearly offers nontrivial predictions for cluster phases, weak
gels and relaxation anomalies. But it neglects two important pieces of
physics which may complicate the experimental situation considerably.
Firstly, we allowed clusters to dissociate by bond breaking, but
neglected any internal restructuring within the clusters.  By allowing
these to coarsen into globules, this could cause both the cluster
fluid and weak gel phases to have finite lifetime.  Moreover, for
$f(N) = 0$ the breaking and rebonding of the ${\cal O}(N)$ dangling
ends within a cluster may cause local restructuring even before the
flow reaches the MCT surface.  This issue is less important in the
presence of a weak, long--range repulsion $f(N)> 0$
(e.g. \cite{yethiraj-blaaderen:2003}), which, like bond--breaking,
drives an upward flow in Fig.~\ref{fig:flow}.  In favorable cases this
could dominate the bond--breaking term at large $\varepsilon$, and
allow the CMCT scenario to establish on a timescale during which local
reconstruction remains negligible.

The second, closely related, omission from the CMCT predictions of
Fig.~\ref{fig:segrelike} is the interplay of cluster formation with
liquid--gas type phase separation \cite{miller-frenkel:2003}. To be
confident of observing the CMCT line directly at small $\phi$, it must
lie outside the binodal. When that condition is not satisfied, we
expect results that depend on quench rates.  For a very slow quench,
the system first demixes locally and then coarsens, until the denser
coexisting phase, whose properties continue to evolve, arrests by a
CMCT--like mechanism (merging with bare MCT at high enough
density). Slow quenches have recently been examined experimentally,
and these ideas may account for a surprising ``bead phase'' seen at
low $\phi$ \cite{sedgwick-etal:tbp}.

\begin{figure}[t]
\begin{center}
\includegraphics[width=\columnwidth]{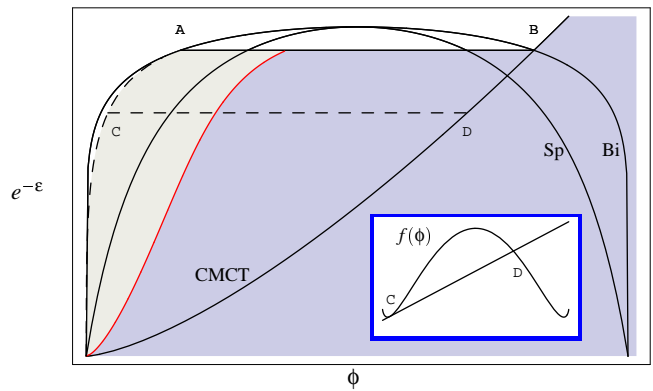}
\end{center}
\caption{Kinetic phase diagram, superimposed on the equilibrium phase
diagram, for an indicative model. The equilibrium free energy density
is that of a lattice gas, $f(\phi) = \phi\ln\phi +
(\phi_m-\phi)\ln(\phi_m-\phi)- \lambda\phi(\phi_m-\phi)$ with $\phi_m
= 0.64$ chosen as a maximum packing fraction. The second virial is
matched to the colloidal system so that $\lambda(\varepsilon) = 1/2
-B_2(\varepsilon) v$ with $v$ the volume of a colloid; $B_2$ is that
of a square well potential of range $\delta =0.1$. The CMCT line is
found for the colloid system as in Fig.~\ref{fig:segrelike}; equilibrium
binodal (Bi) and spinodal (Sp) are marked. \emph{Inset:} modified tangent
construction on $f(\phi)$ \cite{frank-keller:88} in which states beyond D
are deemed inaccessible. The gel line (with percolation threshold set
at phase volume 0.3 of the arrested phase) bounds the darker shaded
region. Between this and the modified binodal (dashed at left) lies
the cluster fluid phase (lighter shading). }
\label{fig:gelgas}
\end{figure}

For faster quench rates, local arrest can precede coarsening so that
at small $\phi$, large fractal clusters will at first evolve,
regardless of the underlying phase behavior.  The resulting scenario
is explored, within an indicative model, in Fig.~\ref{fig:gelgas}.  We
follow Keller \cite{frank-keller:88} in arguing that the densest phase
capable of formation by demixing arises where the glass line
intersects the binodal (point B). The nonergodicity line should depart
from the CMCT prediction at this point, and track to lower densities
along the tie line BA. This describes a locally demixed state in which
only the denser (gel--like) B--component is nonergodic. At high $\phi$
the B--phase percolates giving a finite modulus gel; but at lower
$\phi$ one expects a cluster fluid, now made of disconnected gel--like
fragments. The `gel line', separating states of finite modulus from
those without, corresponds to the geometric percolation threshold {\em
of the arrested phase}. Deeper quenches should behave similarly, with
local coexistence between a dilute fluid C and an arrested state D
whose density lies on the CMCT line. This is governed by a modified
common tangent \cite{frank-keller:88} in which inaccessible densities
(beyond point D) are prohibited (Fig.~\ref{fig:gelgas}, inset)
shifting the binodals inward. Along the tieline CD, the gel line again
corresponds to percolation of the arrested phase. At still deeper
quenches the CMCT line lies within a (renormalized) spinodal curve;
this local instability complicates matters further, and is ignored in
the indicative model of Fig.~\ref{fig:gelgas}. But a transition from
percolating gel to a cluster phase of gel fragments should
remain. Finally, at infinite $\varepsilon$ the CMCT line, the gel
line, the equilibrium spinodal, and the binodal all meet the ICA fixed
point at the origin.  Notably, this scenario allows the `marginal gel'
behavior to extend inside the gel phase: indeed a range of
compositions, including the AB tieline and a region below it, now have
local phase separation involving a marginal gel
(Fig.~\ref{fig:gelgas}).

We now confront the implications of our work with experimental
data. The most direct consequence of our hypothesis of a double
ergodicity breaking, a fractal gel structure, is a very common
observation. It has inspired successful phenomenological gel models,
e.g.~\cite{shih-etal:90,krall-weitz:98,trappe-weitz:2000}, for which
CMCT could provide a more rational (albeit still speculative)
basis. Experiments \cite{dinsmore-weitz:2002} demonstrate that even
for moderate attractions the assumed closeness to the ICA limit is
well justified at low volume fractions in some cases, while in others
the measured fractal dimension lies higher than for the appropriate
ICA model \cite{rojas-etal:2003}.  The latter could well be a
consequence of a partial phase separation followed by arrest, as
discussed above.

Secondly, CMCT predicts dynamic anomalies along and near much of the
gel line (including the part that lies along AB tieline of
Fig.~\ref{fig:gelgas}). In this context, the observation in
Ref.\cite{segre-etal:2001} of a large exponent ($5.5\pm 1$ at
$\phi\approx 0.1$) for the divergence of the terminal time, and of
slow, possibly logarithmic, structural relaxation close to the gel
line at low $\phi$ are both suggestive of CMCT.

A final comparison involves the nonequilibrium phase diagram, where a
striking feature at low concentrations (e.g.\
\cite{poon-etal:99,trappe-weitz:2000,segre-etal:2001}) is indeed the
existence of a phase of large mobile aggregates, lying between the
ordinary fluid phase and the gel. In fact, the phase diagrams in
Fig.~5 of Ref.\cite{segre-etal:2001} and Fig.~2 of
Ref.\cite{verduin-dhont:95} are very similar to our
Figs.~\ref{fig:segrelike},~\ref{fig:gelgas}, respectively.
As mentioned above, an interplay of CMCT with phase separation is
expected for most colloidal parameters, including those of
Ref.\cite{segre-etal:2001}. Fig.~\ref{fig:segrelike} therefore shows
also the phase boundaries adjusted to allow for phase separation
according to the indicative model of Fig.~\ref{fig:gelgas}. In this
representation the cluster/gel boundary remains nearly straight,
although shifted downward somewhat. The results remain qualitatively
consistent with Ref.\cite{segre-etal:2001}; given the indicative
character of the model, this is encouraging.  Since the system is
among those for which a long range Coulomb repulsion might be present
\cite{yethiraj-blaaderen:2003}, it is possible that
$f(N)>0$. Qualitatively, this would move the calculated gel line
upward, stabilizing the cluster phase.

In summary, we have proposed a new framework to describe weak gelation
based on mode--coupling theory applied to clusters (CMCT). This
certainly contains speculative elements but gives nontrivial
predictions, several of which compare well to existing experimental
data.  Underlying our picture of a double ergodicity breaking is the
premise of a time scale separation. Our ``cluster glass'' view of weak
gels is a meaningful concept on time scales longer than the time for
cluster formation, but short compared to typical times over which a
cluster reconstructs as it evolves towards equilibrium. This condition
is more easily satisfied in systems where a repulsion between clusters
builds up as they grow (e.g. weakly charged particles), limiting their
final size. When the size is limited by bond--breaking, the required
time scale separation is less likely.  But even if the resulting weak
gels are not true examples of broken ergodicity and are only temporary
structures, CMCT may offer valuable insights into their
behavior. Further experimentally relevant consequences of the scheme
will be the subject of future work.


We thank Mark Haw, Daan Frenkel, Matthias Fuchs, Helen Sedgwick,
Matthias Sperl and Dave Weitz for valuable discussions, and the EC
for financial support.


\begin{thebibliography}{21}
\bibitem{barrier} It slightly distorts the phase diagram (e.g.\ it
shifts the repulsive glass transition to slightly smaller volume
fractions), and may turn the spinodal into an instability towards
micro--phase separation. These effects are not crucial to the present
discussion.
\end{thebibliography}

\end{document}